# Parallel Gauss-Jordan Elimination and System Reduction for Efficient Circuit Simulation

Filip Noveski [a], Elena Hadzieva [b,*]

[a,b] University of Information Science and Technology "St. Paul the Apostle", Partizanska bb, Ohrid, N. Macedonia,

**Abstract:** For the purposes of electric circuit simulation, we consider an iterative simulation model based on solving systems of linear equations by Gauss-Jordan elimination (GJE) for individual moments in time. To accelerate the simulation, we propose two independent novel approaches: a parallel GJE algorithm and partial system reduction prior to the start of iterations. The former is based on a well-known strategy applied for the first time in this context, whereas the latter, to the best of our knowledge, proposes an entirely new system reduction approach. To evaluate performance, we implement these algorithms in C++ using OpenMP and run them on various input matrices. Our analyses of the individual methods show improved performance, whilst combining them maintains parallel efficiency after partial reduction on medium-sized matrices and even improves efficiency on the largest matrices on the tested machine.

**Keywords:** Electric Circuit, Linear System, Parallel Gauss-Jordan Elimination, System Reduction, Algorithm Efficiency.

**MSC:** 65F05 (Primary), 65Y20, 94C05, 65Y05.

## 1. Introduction

Electric circuits are fundamental to many aspects of modern technology, from consumer electronics and communication networks to medical devices and industrial control systems. Simulating them can be invaluable – not only for enabling engineers to predict circuit behaviour, optimise performance, and reduce development costs, but also for helping students visualise circuit characteristics and laws, reinforcing their theoretical knowledge in ways similar to working with physical circuits.

In this paper, we analyse electric circuits consisting of $n$ branches. By applying Kirchhoff's Current Law (KCL) and Kirchhoff's Voltage Law (KVL), the following system of $n$ linearly independent equations in $n$ unknowns is obtained (throughout this paper, elements and matrix rows/columns are indexed starting from 0):

$$\begin{cases} a_{0,0}x_0 + a_{0,1}x_1 + \cdots + a_{0,n-1}x_{n-1} = b_0 \\ a_{1,0}x_0 + a_{1,1}x_1 + \cdots + a_{1,n-1}x_{n-1} = b_1 \\ \quad\quad\quad\quad\quad \vdots \\ a_{n-1,0}x_0 + a_{n-1,1}x_1 + \cdots + a_{n-1,n-1}x_{n-1} = b_{n-1} \end{cases} \quad (1.1)$$

Depending on the structure of the circuit, each unknown $x_i$ may either be the current $I_i$ flowing through the branch with the index $i$ or the electromotive force (EMF) $E_i$ that the source of the electric current in the branch $i$ must provide to maintain the specified magnitude of current.

*corresponding author: elena.hadzieva@uist.edu.mk

The corresponding augmented matrix for the system (1.1) is of dimension $n \times (n+1)$ and of rank $n$:

$$\begin{bmatrix} a_{0,0} & a_{0,1} & \cdots & a_{0,n-1} & \vdots & b_0 \\ a_{1,0} & a_{1,1} & \cdots & a_{1,n-1} & \vdots & b_1 \\ \vdots & \vdots & \ddots & \vdots & \vdots & \vdots \\ a_{n-1,0} & a_{n-1,1} & \cdots & a_{n-1,n-1} & \vdots & b_{n-1} \end{bmatrix}_{n \times (n+1)} \quad (1.2)$$

By applying Gauss-Jordan elimination (GJE) to the augmented matrix (1.2), the unique solution for each unknown is obtained [20], allowing the circuit state to be determined. When discussing GJE in the following sections, we refer to the $i$-th row and column as $R_i$ and $C_i$, respectively. In the case where the circuit has components with parameters that change over time, the parameter value at each moment $t_i$ can be injected into the augmented matrix before applying GJE. By performing this operation on predefined time intervals defined by a constant time step $\Delta t$, the circuit can be simulated iteratively, with the rate at which the system is solved impacting the value of $\Delta t$ and thus the simulation accuracy.

The paper is organised as follows. Section 2 provides an overview of related work, with particular emphasis on matrix solvers and approaches to parallelisation. Section 3 presents the main contributions of the paper. We first explore parallel programming as a means to speed up the GJE algorithm. Next, we turn towards partially solving the system before initiating iterations, thereby reducing the number of operations per iteration. Section 4 contains the experimental results and discussion, including evaluations on randomly generated matrices and circuit-based systems, followed by a summary and comparison of the individual findings. The concluding remarks are in the last, fifth section.

## 2. Related Articles

In light of the many use cases for Gauss-Jordan elimination (GJE), numerous papers have explored ways to parallelise the algorithm, both on multicore CPUs and by using GPUs as accelerators. Catalán et al. [18] have proposed task-parallel GJE algorithms targeting multicore CPUs for the inversion of dense matrices. Their approach is to start by partitioning the matrix into column blocks, which are further partitioned into either row blocks or thinner column blocks, thereby recording a performance improvement using OpenMP.

Alassafi and Alsenani [11] likewise explored parallel GJE for matrix inversion, targeting multicore CPUs with OpenMP. They utilised LU decomposition, splitting the procedure into multiple independent tasks, enabling straightforward parallelisation and achieving positive parallel efficiency, albeit diminishing as the level of parallelisation and matrix size increase (both individually and in combination).

Lu et al. [2] investigated GJE (among other algorithms) using MPI to invert very large matrices. They also discovered a reduction in execution time as the number of processors increased, but parallel efficiency declined. Another paper by Catalán et al. [19] expands upon their work in the reference [18] with further testing. Here, they found that their advanced task-parallel GJE algorithm is consistently more performant than an Intel MKL counterpart they also tested.

Atasoy et al. [12] have proposed a GPU-based parallel GJE algorithm in the context of simulating an electric circuit using Nvidia's Compute Unified Device Architecture (CUDA), achieving a notable improvement in performance.

In [4], the authors study the Gaussian Elimination method, and Parallel Implicit Elimination Methods and their effects on the performance of processing. According to their findings, OpenMP is, on average, superior to Pthreads and MPI.

Marrakchi and Kaaniche in [21] use the Gaussian Elimination method to solve band diagonally dominant linear systems by allocating balanced tasks to suitable cores, which are performed parallelly. Later, in [22], these authors review modern approaches to solving sparse triangular linear systems (spTRSV) on parallel and distributed architectures. Liu et al. [23] propose synchronization-free parallel algorithms for sparse triangular solve kernels, SpTRSV and SpTRSM, in which optimisation paths are adaptively selected for best parallelism in the case of multiple right-hand sides. Parallel implementations focused on solving large tridiagonal systems on multi-core architectures, based on combining the parallel PCR and Thomas algorithms are presented in [16].

In [15], the authors present a shared-memory parallel GJE algorithm that utilises pivot row streaming to reduce synchronisation between the processors. HOGWILD! [6] and CYCLADES [24] are both examples of algorithms that aim to reduce synchronisation during Stochastic Gradient Descent (SGD) through probabilistic correctness and careful scheduling, respectively. NOMAD [7] uses local copies to achieve the same effect of reduced synchronisation.

System reduction is a well-established approach for improving computational efficiency and reducing overall computational complexity, and it has been widely studied across different classes of mathematical systems. Partial and total reduction techniques for linear systems of operator equations have been investigated in [3, 8, 9], local reduction of a decision system has been investigated in [5], while the reduction of systems of differential and difference equations can be observed in [14] and [17]. Jenko [10] explores simulating differential equations that describe electric circuits for educational purposes, elaborating that the approach can aid the students' understanding in ways similar to building physical circuits.

In this work, we present a parallel GJE algorithm that splits the matrix into batches of contiguous rows, each assigned to an individual processor. To minimise the need for synchronisation and improve efficiency, a separate pivot matrix is introduced that stores pivot rows in a read-only format, which are appended as they become ready. Although this is a well-established strategy for efficient shared-memory parallel algorithms, our approach differs from those in the aforementioned papers in that we store all shared information throughout the progression of the algorithm. We then explore a problem-specific approach of partial system reduction. This approach utilises the characteristics of systems of linear equations with time-variable entries, aiming to lower the order of the augmented matrix, thereby improving performance in discrete-time iterations. Finally, we evaluate the performance of these two approaches in isolation and in combination.

To the best of our knowledge, these specific forms of parallelisation of GJE and partial reduction – applied in this manner and for this computational setting – have not been previously reported in the literature.

## 3. Parallelisation of the Gauss-Jordan Elimination Method and Reduction of the Linear System

To optimise the process of time-dependent iterative application of GJE, we explore two independent approaches. The first entails harnessing multiple processors to reduce the execution time. The second focuses on reducing a part of the matrix before initiating time-based iterations, aiming to reduce the number of operations per iteration.

### 3.1. Gauss-Jordan Elimination

In this section, we first define a baseline sequential GJE algorithm that can be used to solve the system $(1.2)$, outlining methods to optimise the process before parallelising it. We then introduce a row-wise parallel GJE algorithm that splits the work across multiple CPU cores via block partitioning.

#### 3.1.1. Baseline Sequential Algorithm

In line with Gauss-Jordan elimination (GJE), we use an iterative approach whereby each iteration is centred around a pivot element $a_{i,i}, i \in \{0, 1, 2, \ldots, n-1\}$, progressing in order. Each iteration applies row-equivalent transformations such that the first iteration $i = 0$ brings the augmented matrix $(1.2)$ to the following form:

$$\begin{bmatrix} 1 & a'_{0,1} & \cdots & a'_{0,n-1} & \vdots & b'_0 \\ 0 & a'_{1,1} & \cdots & a'_{1,n-1} & \vdots & b'_1 \\ \vdots & \vdots & \ddots & \vdots & \vdots & \vdots \\ 0 & a'_{n-1,1} & \cdots & a'_{n-1,n-1} & \vdots & b'_{n-1} \end{bmatrix}_{n \times (n+1)} \quad (3.1)$$

As no further changes will occur in the leftmost column $C_0$ throughout the remaining iterations, the algorithm will continue to work on the right-hand $n \times n$ submatrix, shown in (3.2), to reduce the number of operations required.

$$\begin{bmatrix} a'_{0,1} & a'_{0,2} & \cdots & a'_{0,n-1} & \vdots & b'_0 \\ a'_{1,1} & a'_{1,2} & \cdots & a'_{1,n-1} & \vdots & b'_1 \\ \vdots & \vdots & \ddots & \vdots & \vdots & \vdots \\ a'_{n-1,1} & a'_{n-1,2} & \cdots & a'_{n-1,n-1} & \vdots & b'_{n-1} \end{bmatrix}_{n \times n} \quad (3.2)$$

The second iteration for $i = 1$ focuses on the second column of the complete matrix (3.1). The elementary row-operations are applied as required to bring this column to the state: $[0 \quad 1 \quad 0 \quad 0 \quad \cdots \quad 0]^T$.

The final iteration, for $i = n - 1$, will only work with the rightmost two columns $C_{n-1}$ and $C_n$ shown in (3.3). Following this iteration, the submatrix of free terms, that is $C_n$, shown in (3.4), contains the solution to each unknown, whilst the left-hand $n \times n$ submatrix contains the identity matrix.

$$\begin{bmatrix} a''_{0,n-1} & \vdots & b''_0 \\ a''_{1,n-1} & \vdots & b''_1 \\ \vdots & \vdots & \vdots \\ a''_{n-1,n-1} & \vdots & b''_{n-1} \end{bmatrix}_{n \times 2} \tag{3.3}$$

$$\begin{bmatrix} x_0 \\ x_1 \\ \vdots \\ x_{n-1} \end{bmatrix}_{n \times 1} \tag{3.4}$$

To further optimise this algorithm, if the pivot column already contains a zero in a target row, we skip a row-addition operation to that row. This procedure is shown in Algorithm 1.

```
 1 Function GaussJordanSerial(Matrix A, integer n)
 2     For i := 0 to n - 1 do
 3         If (aᵢ,ᵢ = 0) then
 4             Rᵢ ↔ Rⱼ where aⱼ,ᵢ ≠ 0 and j > i
 5         EndIf
 6
 7         Iterate(A, n, i)
 8     EndFor
 9 EndFunction
10
11 Function Iterate(Matrix A, integer n, integer i)
12     m = 1 / aᵢ,ᵢ
13     If (m ≠ 1)
14         mRᵢ → Rᵢ
15     EndIf
16
17     For j := 0 to n - 1 do
18         If (j = i) then
19             skip
20         EndIf
21
22         If (aⱼ,ᵢ ≠ 0) then
23             Rⱼ + aⱼ,ᵢRᵢ → Rⱼ
24         EndIf
25     EndFor
26 EndFunction
```

**Algorithm 1.** Sequential Gauss-Jordan elimination

In the following text, we provide an example to illustrate the progression of Algorithm 1. For this example, we use the circuit shown in Figure 1 that includes the variable resistor $R_2$, allowing us to showcase the iterative simulation model. By applying KCL and KVL to this circuit, we form the following augmented matrix:

$$\begin{bmatrix} 1 & -1 & -1 & \vdots & 0 \\ 600 & 900 & 0 & \vdots & 12 \\ 0 & -900 & R_2(t) & \vdots & 0 \end{bmatrix}_{3 \times 4} \tag{3.5}$$

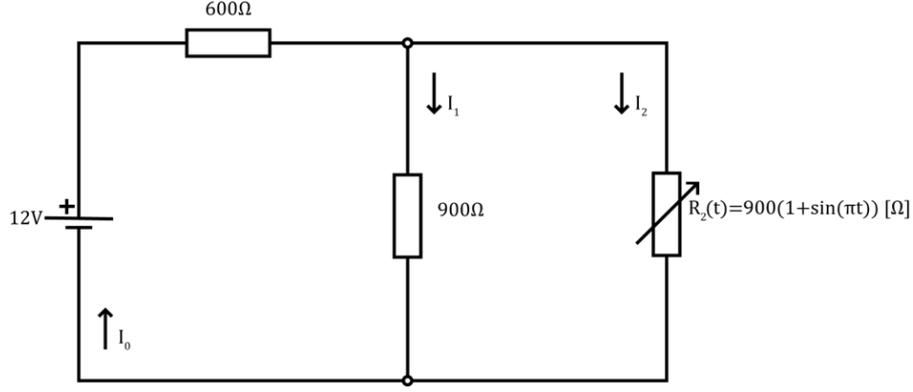

**Figure 1.** Example circuit with 3 branches and 2 nodes

Before applying GJE, we must insert a value for $R_2(t)$, for which we need a moment in time $t$. For this example, we consider the moment $t = 0.5s = 500ms$, which gives us a resistance $R_2(0.5) = 900 \cdot (1 + \sin(0.5\pi)) = 1800\Omega$. Upon injecting this value into the system (3.5), we can apply Algorithm 1 to find the following solution:

$$\begin{bmatrix} 1 & -1 & -1 & : & 0 \\ 600 & 900 & 0 & : & 12 \\ 0 & -900 & 1800 & : & 0 \end{bmatrix} \rightarrow \begin{bmatrix} 1 & 0 & 0 & : & 0.01 \\ 0 & 1 & 0 & : & 0.00667 \\ 0 & 0 & 1 & : & 0.00333 \end{bmatrix} \quad (3.6)$$

The above procedure can be used to simulate electric circuits by considering adequate moments in time. To simulate circuits iteratively with uniform sample intervals, the simulator would start at an initial moment $t_0 = 0s$ and advance by the constant time step $\Delta t = t_{i+1} - t_i$, evaluating the state for each moment $t_i$.

### 3.1.2. Parallelising the Algorithm

To parallelise the GJE algorithm on a multicore CPU, we split the matrix into blocks of consecutive rows of equal size, such that $b = n/p$, where $n$ is the number of rows and $p$ is the number of processors running the operation. For simplicity, in the following text, we assume that $n$ is divisible by $p$. The processor $p_i \in \{0, 1, 2, \ldots, p-1\}$ is assigned rows $B_i = \{R_{bp_i}, R_{bp_i+1}, R_{bp_i+2}, \ldots, R_{bp_{i+1}-1}\}$.

While most operations allow each processor $p_i$ to freely modify the rows in $B_i$ without synchronisation with the other processors, if another processor $p_j$ needs to read its relevant pivot row, conflicts may arise. To tackle this, we propose using a separate pivot matrix $P$, shown in (3.7)[1], that stores all prepared pivot rows and allows all processors to read them without synchronisation. Entries marked with an en-dash (−) are not allocated as their value is known to be zero and is ignored during the iteration involving the corresponding pivot row. Inherently, the number of elements, $|P|$, this pivot matrix has for an augmented matrix with $n$ rows is evaluated by the formula (3.8).

---

[1] When we use $p_{i,j}$ with two indices, we refer to an entry in the pivot matrix $P$. If we provide only one index in the subscript ($p_i$), we refer to a processor.

$$\begin{bmatrix} p_{0,0} & p_{0,1} & p_{0,2} & \cdots & p_{0,n-2} & \vdots & p_{0,n-1} \\ - & p_{1,1} & p_{1,2} & \cdots & p_{1,n-2} & \vdots & p_{1,n-1} \\ - & - & p_{2,2} & \cdots & p_{2,n-2} & \vdots & p_{2,n-1} \\ \vdots & \vdots & \vdots & \ddots & \vdots & \vdots & \vdots \\ - & - & - & \cdots & p_{n-2,n-2} & \vdots & p_{n-2,n-1} \\ - & - & - & \cdots & - & \vdots & - \end{bmatrix} \qquad (3.7)$$

$$|P| = \frac{(n-1)\cdot(n+2)}{2} \qquad (3.8)$$

The procedure of the parallel GJE algorithm is shown in Algorithm 2.

### 3.1.3. Time Complexity Analysis of Parallel Gauss-Jordan Elimination

The following text analyses the time complexity of the parallel Gauss-Jordan elimination algorithm shown in Algorithm 2. The analysis is performed over an $n \times (n+1)$ matrix provided as the input with $p$ processors applying GJE over it. We also assume that $n$ is divisible by $p$, resulting in an even split of rows across all processors.

Algorithm 2 shows that, before forking threads for parallel execution, a single thread checks the validity of the first pivot element, swapping rows if necessary, and appends the first pivot row $R_0$ to the pivot matrix $P$. We denote the time complexity of this phase with $T_s(n)$. As both of these operations display linear growth, we express $T_s(n)$ as:

$$T_s(n) = n \qquad (3.9)$$

The parallel region, described by the function GaussJordanThread, is split into three phases: pre-diagonal, diagonal, and post-diagonal. For this calculation, we denote the rank of an arbitrary thread with $r$. The time complexity of the parallel region is denoted by $T_p(n, p)$ and evaluated as follows:

$$T_p(n,p) = \sum_{i=0}^{\frac{rn}{p}-1} \frac{n}{p}(n-i+1) + \sum_{i=\frac{rn}{p}}^{\frac{(r+1)n}{p}-1} \frac{n}{p}(n-i+1) + \sum_{i=\frac{(r+1)n}{p}}^{n-1} \frac{n}{p}(n-i+1) = \frac{n^3 + 3n^2}{2p} \qquad (3.10)$$

In the time complexity evaluation (3.10), each sum corresponds to its respective phase, namely the pre-diagonal, diagonal and post-diagonal phases. The first and last terms result from each thread working on $n/p$ rows and only working on the rightmost $n - i + 1$ entries in each row during the iteration with index $i$. The middle term is a sum of the time complexities of the integrated Gaussian elimination and back-substitution steps. During each iteration with index $i$, the Gaussian elimination step involves dividing the pivot row by the pivot element and appending it to the pivot matrix $P$ (1 row-operation) and adding a multiple of it to the $n/p - (i \bmod n/p) - 1$ rows beneath it. Likewise, the back-substitution phase requires a multiple of the pivot row to be added to the $i \bmod n/p$ rows above it. Combining these steps, we conclude that $n/p$ rows are involved, with $n - i + 1$ elements operated on during the iteration with the index $i$, hence the sum present in the middle in the evaluation (3.10).

```
 1  Function GaussJordanParallel(Matrix A, integer n, integer p)
 2      P := PivotMatrix(n)
 3      If (a_{0,0} = 0) then
 4          R_0 ↔ R_i where a_{i,0} ≠ 0 and i > 0
 5      EndIf
 6      Append R_0 / a_{0,0} to P
 7      Parallel in p threads
 8          GaussJordanThread(A, n, P)
 9      EndParallel
10  EndFunction
11
12  Function GaussJordanThread(Matrix A, integer n, PivotMatrix P)
13      r := getThreadRank()
14      b := n / p
15      ds := r * b
16      de := (r + 1) * b – 1
17
18      (* pre-diagonal *)
19      For i := 0 to ds - 1 do
20          While (R_i not in P) do
21              wait
22          EndWhile
23          For j := ds to de – 1 do
24              (* below, R_i is read from P; R_j is modified in A *)
25              R_j + a_{j,i}R_i → R_j
26              If (a_{j,i+1} ≠ 0) then
27                  label R_j as valid pivot row
28              EndIf
29          EndFor
30      EndFor
31
32      (* diagonal *)
33      For i := ds to de - 1 do
34          Apply Gaussian elimination
35              ↳ Append R_j / a_{j,j} to P as soon as row op is applied
36              ↳ Look to other threads and local rows if swap needed
37          Apply back-substitution
38      EndFor
39
40      (* post-diagonal *)
41      For i := de to n - 1 do
42          While (R_i not in P) do
43              wait
44          EndWhile
45          For j := ds to de – 1 do
46              (* below, R_i is read from P; R_j is modified in A *)
47              R_j + a_{j,i}R_i → R_j
48          EndFor
49      EndFor
50  EndFunction
```

**Algorithm 2**. Parallel Gauss-Jordan elimination

From the time complexities (3.9) and (3.10), we conclude that the algorithm has a parallel time complexity per processor of $O\left(n^3/p + n\right)$, indicating that the cubic growth is split across the $p$ processors, leaving a linear term for the sequential overhead. Furthermore, as the rank of the thread is not present in the result for (3.10), we can assume that work is split evenly across the processors.

### 3.2. Partial Reduction of Linear Systems

In this section, we present the algorithm for partial reduction of the system of linear equations prior to the initiation of the iterative phase. We first outline the requirements for this operation to be valid, then analyse and describe the row operations that can be applied. We close out the section with a description of the algorithm itself.

#### 3.2.1. Requirements

In the augmented matrix, we represent each entry $a_{ij}$ as a sum of two terms:

$$a_{ij} = c_{ij} + v_{ij} = \begin{cases} c_{ij}, & \text{if } v_{ij} = 0, \\ c_{ij} + v_{ij}, & \text{if } v_{ij} = f_{ij}(t), \end{cases} \quad (3.11)$$

In memory, we assume that the base matrix stores only the $c_{ij}$ term of each entry, regardless of whether a non-zero $v_{ij}$ term is assigned to it. During each iteration, the adequate $v_{ij} = f_{ij}(t)$ terms are evaluated, and whole units are added to the appropriate entries. Finally, the implemented Gauss-Jordan elimination function is not aware of any $v_{ij}$ terms that are to be added to the matrix.

Based on these assumptions, after any row operation that results in the element $a'_{ij}$ being present at a given index, we apply the following restrictions on the partial reduction algorithm:

1. $a'_{ij}$ must maintain a whole $v_{ij}$ term in it, i.e. $a'_{ij} = c'_{ij} + v_{ij}$.
2. A new variable term $v_{kl}$ not present in $a_{ij}$ must not be introduced to $a'_{ij}$.

#### 3.2.2. Description of Valid Row-Operations

To find the row operations that are valid, we analyse each one based on how it is used and the resulting impact on each element. We say that valid operations are those that obey both criteria introduced in Subsection 3.2.1.

i) Multiplying a row by a non-zero number

During the application of GJE, the row-multiplication operation is applied to divide the pivot row by the pivot element (see Section 3.1):

$$\frac{1}{a_{ii}} \cdot R_i \to R_i \quad \Rightarrow \quad a'_{ij} = \frac{a_{ij}}{a_{ii}} \quad (3.12)$$

By analysing each case for $a_{ij}$ and $a_{ii}$, we get the following results:

$$a'_{ij} = \begin{cases} \dfrac{1}{c_{ii}} \cdot c_{ij}, & a_{ij} = c_{ij} \text{ and } a_{ii} = c_{ii} \\ \dfrac{1}{c_{ii}} \cdot c_{ij} + \dfrac{1}{c_{ii}} \cdot v_{ij}, & a_{ij} = c_{ij} + v_{ij} \text{ and } a_{ii} = c_{ii} \\ \dfrac{1}{c_{ii} + v_{ii}} \cdot c_{ij}, & a_{ij} = c_{ij} \text{ and } a_{ii} = c_{ii} + v_{ii} \\ \dfrac{c_{ij}}{c_{ii} + v_{ii}} + \dfrac{v_{ij}}{c_{ii} + v_{ii}}, & a_{ij} = c_{ij} + v_{ij} \text{ and } a_{ii} = c_{ii} + v_{ii} \end{cases} \quad (3.13)$$

From the cases (3.13), we conclude that (3.12) is only valid if both $a_{ij}$ and $a_{ii}$ are constant entries. However, if $\exists a_{ij} \in R_i, a_{ij} = c_{ij} + v_{ij}, v_{ij} \neq 0$, (3.8) shows that $v_{ij}$ is divided by $a_{ii}$ after the operation, violating the first criterion. Likewise, if $a_{ii} = c_{ii} + v_{ii}, v_{ii} \neq 0$, the second criterion is violated as $v_{ii}$ is introduced to the other entries.

ii) Adding a multiple of one row to another

To make every element in the pivot column except the pivot element equal to zero, a multiple of the pivot row is added to each of the remaining rows by the following rule:

$$-a_{ki} \cdot R_i + R_k \to R_k \Rightarrow a'_{kj} = -a_{ki} \cdot a_{ij} + a_{kj} \quad (3.14)$$

By considering the possible values of $a_{ki}$, $a_{ij}$ and $a_{kj}$, we evaluate the following possible results for $a'_{kj}$:

$$a'_{kj} = \begin{cases} -a_{ki} \cdot a_{ij} + c_{kj} + v_{kj}, & a_{kj} = c_{kj} + v_{kj} \\ -a_{ki}(c_{ij} + v_{ij}) + a_{kj}, & a_{ij} = c_{ij} + v_{ij} \\ -a_{ij}(c_{ki} + v_{ki}) + a_{kj}, & a_{ki} = c_{ki} + v_{ki} \end{cases} \quad (3.15)$$

The cases (3.15) show that the validity of (3.14) does not depend on the value contained in $a_{kj}$. But, if $\exists a_{ij} \in R_i, a_{ij} = c_{ij} + v_{ij}, v_{ij} \neq 0$ or $a_{ki} = c_{ki} + v_{ki}, v_{ki} \neq 0$, the second criterion is violated. Inherently, we apply (3.9) if neither the pivot row nor the pivot column contains elements with variable terms.

iii) Swapping rows

The final operation to consider is the row-swapping one, impacting the elements like so:

$$R_i \leftrightarrow R_k \Rightarrow a'_{ij} = a_{kj} \text{ and } a'_{kj} = a_{ij} \quad (3.16)$$

In this case, if neither row contains dynamic elements, the operation is valid. But, if either $a_{ij}$ or $a_{kj}$ has a dynamic term, following the row swap, one's dynamic term is introduced to the other and lost in the former, violating both criteria. Hence, a swap involving a dynamic row is not legal.

### 3.2.3. Partial Reduction Algorithm

When initiating partial reduction of the system, we ensure that the correct row-equivalent transformations, as discussed in Subsection 3.2.2, are applied by defining a boundary $\beta$ past which pivot elements cannot be selected. If $C_i$ is the first column with an entry $a_{ki} = c_{ki} + v_{ki}$ and $R_j$ is the first row with an entry $a_{jh} = c_{jh} + v_{jh}$, then the boundary $\beta = \min(i, j)$. From here, the only pivot elements that the partial algorithm selects are $a_{i,i}, i \in \{0, 1, 2, \dots, \beta - 1\}$. The procedure describing this algorithm is shown in Algorithm 3.

```
1 Function GaussJordanPartial(Matrix A, integer n, integer β) : integer
2     For i := 0 to β - 1 do
3         If (aᵢ,ᵢ = 0) then
4             Rᵢ ↔ Rⱼ where aⱼ,ᵢ != 0 and j > i and j < β
5                 ↳ if not possible, return i
6         EndIf
7
8         Iterate(A, n, i)
9     EndFor
10
11    Return β
12 EndFunction
```

**Algorithm 3**. Partial Gauss-Jordan elimination. The Iterate function is identical to the one shown in Algorithm 1.

Following this operation, all variable $v_{ij}$ terms may be added to the relevant entries in the augmented matrix as before without violating the system's correctness. To solve the resulting system, either one of the two GJE algorithms discussed in Section 3.1 may be used.

In the following text, we demonstrate the application of Algorithm 3 over the circuit shown in Figure 1 and its corresponding system (3.5). From (3.5), we can observe that the only variable term $R_2(t)$ is to be injected at the index $(3, 2)$. Hence, we evaluate the boundary $\beta = \min(3, 2) = 2$. From here, the application of Algorithm 3 over the system (3.5) results in the following partially reduced augmented matrix (note that in this example, we display $R_2(t)$ in the augmented matrix to show that both criteria elaborated in Subsection 3.2.1 are obeyed upon completion; however, as discussed in the above subsection, we assume that the machine does not store this term in memory at this time):

$$\begin{bmatrix} 1 & -1 & -1 & \vdots & 0 \\ 600 & 900 & 0 & \vdots & 12 \\ 0 & -900 & R_2(t) & \vdots & 0 \end{bmatrix} \xrightarrow{R_2 - 600R_1 \to R_2} \begin{bmatrix} 1 & -1 & -1 & \vdots & 0 \\ 0 & 1500 & 600 & \vdots & 12 \\ 0 & -900 & R_2(t) & \vdots & 0 \end{bmatrix} \quad (3.17)$$

$$\xrightarrow{\frac{1}{1500} R_2 \to R_2} \begin{bmatrix} 1 & -1 & -1 & \vdots & 0 \\ 0 & 1 & 0.4 & \vdots & 0.008 \\ 0 & -900 & R_2(t) & \vdots & 0 \end{bmatrix} \xrightarrow{\substack{R_1 + R_2 \to R_1 \\ R_3 + 900R_2 \to R_3}} \begin{bmatrix} 1 & 0 & -0.6 & \vdots & 0.008 \\ 0 & 1 & 0.4 & \vdots & 0.008 \\ 0 & 0 & 360 + R_2(t) & \vdots & 7.2 \end{bmatrix}$$

From the result obtained in **Error! Reference source not found.**, we observe that, although the constant of the entry at index $(2, 2)$ has changed from 0 to 360, the entry still maintains a whole $R_2(t)$ term and this $R_2(t)$ term has not been introduced to any other entries.

Using the result from (3.17), we can again solve the system for any moment in time $t$. If we select $t = 0.5s$ (same as in the example in Subsection 3.1.1), we evaluate $R_2(t) = 1800\Omega$ and add that value to the entry at index $(2, 2)$. From here, we solve the system like so:

$$\begin{bmatrix} 1 & 0 & -0.6 & \vdots & 0.008 \\ 0 & 1 & 0.4 & \vdots & 0.008 \\ 0 & 0 & 360 + 1800 & \vdots & 7.2 \end{bmatrix} \to \begin{bmatrix} 1 & 0 & 0 & \vdots & 0.01 \\ 0 & 1 & 0 & \vdots & 0.00667 \\ 0 & 0 & 1 & \vdots & 0.00333 \end{bmatrix} \quad (3.18)$$

One may observe that the result attained in (3.18) is equivalent to the one in (3.6) where the entire system is solved at once. However, in this example, fewer row-operations are applied for the iteration for $t = 0.5s$.

### 3.2.4. Time Complexity Analysis of Gauss-Jordan Elimination After Partial System Reduction

In the following text, we analyse the time complexity of the GJE algorithm after the system has been partially reduced. We analyse this operation in terms of an $n \times (n + 1)$ matrix provided as the input, with $r$ columns due to be solved. If the partial system reduction algorithm completed $\beta$ iterations, then $r = n - \beta$.

Based on these parameters, if the baseline sequential GJE algorithm described in Subsection 3.1.1 is used, we evaluate the time complexity as follows:

$$T_{pr}(n, r) = \sum_{i=n-r}^{n-1} (n^2 - n \cdot i + n) = \frac{n \cdot r^2 + 3n \cdot r}{2} \quad (3.19)$$

From the time complexity (3.19), we conclude that the algorithm has a Big O complexity of $O(n \cdot r^2)$. This implies a quadratic decrease in time from partial system reduction based on the number of columns solved. If the parallel GJE algorithm, described in Subsection 3.1.2, is used to solve the remainder of the system, the time complexity is expressed as $O\left(n \cdot r^2 / p\right)$. This allows us to draw the same conclusion regarding time decrease, whilst also observing that the workload balance between the processors is maintained. Although we omit this analysis for brevity, the interested reader may derive the result from the analysis in Subsection 3.1.3.

## 4. Results and Discussion

In this section, we discuss the performance of the algorithms proposed in Section 3. For this purpose, we implement the algorithms using C++ and OpenMP 2.0 [13], and run them on predetermined inputs of different types. We start with randomly generated matrices, which represent dense systems, followed by circuit-based matrices that are sparser. We conclude by comparing the results between these inputs to analyse the performance of the algorithms in each case.

The tests were run on a laptop computer with an Intel Core i7-4700HQ CPU with four physical cores, each supporting SMT. Its base clock frequency is 2.4 GHz with a maximum turbo frequency of 3.4 GHz. Memory is represented by 3 GB of DDR3 RAM. The machine is running Windows 10 with the code being written and built in Microsoft Visual Studio 2022 (with MSVC v143). To run the tests, the C++ project is built as a dynamic-link library (DLL), and it is invoked by a C# program using the BenchmarkDotNet package [1] to run the tests. Each input is run 200 times, with outliers being automatically removed by the aforementioned package.

### 4.1. Gauss-Jordan Elimination on a Randomly Generated Matrix

We open the performance evaluation discussion with tests involving randomly generated matrices. These tests demonstrate the performance on dense systems where many row-addition operations are required and work between the threads can be split more evenly. Our first analysis involves the parallel GJE algorithm described in Subsection 3.1.2, with the results presented in Table 1.

**Table 1.** Recorded times for the sequential and parallel GJE algorithms on randomly generated matrices

| Matrix Size | 1 thread | 2 threads | 4 threads | 8 threads |
|---|---:|---:|---:|---:|
| 128×129 | 437.6 μs | 239.6 μs | 213.6 μs | 124.1 μs |
| 512×513 | 29.4 ms | 15.4 ms | 10.5 ms | 8.1 ms |
| 1024×1025 | 323.1 ms | 171.1 ms | 144.6 ms | 129.7 ms |
| 2040×2041 | 4.3 s | 3.5 s | 3.3 s | 3.5 s |

From Table 1, we can see that parallel performance is generally better than sequential performance across the presented matrix sizes. Peak performance is achieved for the $512 \times 513$ matrix size, with a $3.6 \times$ speedup for the eight-threaded function, with the $128 \times 129$ matrix being solved with comparable performance on the same number of threads. By comparison, on the largest matrices, parallel performance becomes progressively lower as the size increases.

We next turn our discussion to partial system reduction, described in Section 3.2, with the results of the finishing phase (after initial reduction) shown in Table 2.

**Table 2.** Recorded times for the full and partial GJE algorithm executed sequentially on randomly generated matrices

| Matrix Size | No reduction | 19% reduced | 50% reduced |
|---|---:|---:|---:|
| 128×129 | 437.6 μs | 298.5 μs | 130.4 μs |
| 512×513 | 29.4 ms | 19.8 ms | 8.3 ms |
| 1024×1025 | 323.1 ms | 211.3 ms | 90.0 ms |
| 2040×2041 | 4.3 s | 2.8 s | 1.0 s |

These results show a significant improvement in performance from prior partial reduction, with the percentage of time saved by solving less of the matrix being greater than the percentage of columns reduced prior. Furthermore, this advantage is greater on the larger matrices relative to the smaller ones. Both of these conclusions are in line with the time complexity presented in Subsection 3.2.4.

Finally, we analyse the impact of parallelisation on the partially reduced matrix. We only present the results with 50% of the columns having been reduced prior as we expect this to have a greater impact on parallel performance than the smaller reduction amount. The results from these tests are shown in Table 3.

**Table 3.** Recorded times for the partial GJE algorithm with 50% of columns reduced using serial and parallel execution on randomly generated matrices

| Matrix Size | 1 thread | 2 threads | 4 threads | 8 threads |
|---|---|---|---|---|
| **128×129** | 130.4 μs | 79.6 μs | 68.7 μs | 50.5 μs |
| **512×513** | 8.3 ms | 4.7 ms | 3.8 ms | 3.5 ms |
| **1024×1025** | 90.0 ms | 43.8 ms | 42.8 ms | 29.7 ms |
| **2040×2041** | 1 000.3 ms | 654.6 ms | 579.6 ms | 537.9 ms |

Across all matrix sizes, we can observe a speedup from using parallelism. Compared to solving the entire matrix (see Table 1), the two smaller matrices recorded a smaller speedup (2.4 × for the $512 \times 513$ matrix), whilst the larger ones show improved performance with the $1024 \times 1025$ matrix recording a 3 × speedup on eight threads for the partially reduced matrix (comapred to 2.5 × on the entire matrix). Overall, we can conclude that the two approaches can work well together, but their impact on one another depends on the size of the input matrix.

### 4.2. Gauss-Jordan Elimination on a Circuit-Based System

The following text focuses on the performance of the GJE algorithms on circuit-based systems. These systems are sparser, with many entries equal to zero, leading to a less uniform distribution of work between the processors. The results for the parallel GJE algorithm are shown in Table 4.

**Table 4.** Recorded times for the sequential and parallel GJE algorithms on circuit-based matrices

| Matrix Size | 1 thread | 2 threads | 4 threads | 8 threads |
|---|---|---|---|---|
| **128×129** | 200.5 μs | 183.0 μs | 141.0 μs | 312.2 μs |
| **512×513** | 11.1 ms | 9.9 ms | 6.7 ms | 9.8 ms |
| **1024×1025** | 89.1 ms | 74.4 ms | 51.2 ms | 44.1 ms |
| **2040×2041** | 1 078.3 ms | 870.0 ms | 692.8 ms | 435.7 ms |

In contrast to the results shown in Table 1, these tests show the parallel algorithm becoming progressively more efficient as the matrix size increases. Peak speedup occurs at the largest matrix with 2.5 × on eight threads. However, two outliers occur for the smaller two matrices when the same level of parallelism is applied. The median times for these runs are $110.5 \mu s$ and $6.3 ms$ for the $128 \times 129$ and $512 \times 513$ matrices, respectively, with both recording higher standard deviation than the other tests. These insights imply that this setup can be performant for both of these input sizes, however it can also be slowed down by scheduling.

Another observation from these results is that they are notably quicker than those presented in Table 1 for the random matrices. This can be expected due to the sparser nature of the circuit-based matrices.

We now analyse partial system reduction on these matrices, with results presented in Table 5.

Table 5. Recorded times for the full and partial GJE algorithm executed sequentially on circuit-based matrices

| Matrix Size | No reduction | 19% reduced | 50% reduced |
|---|---|---|---|
| 128×129 | 200.5 μs | 79.4 μs | 38.4 μs |
| 512×513 | 11.1 ms | 2.1 ms | 0.5 ms |
| 1024×1025 | 89.1 ms | 17.7 ms | 3.2 ms |
| 2040×2041 | 1 078.3 ms | 176.0 ms | 33.9 ms |

Relative to the runs on the randomly generated matrices, these tests show a notably greater reduction in time following partial reduction. With 19% reduction, the savings in time range from 60% on the smallest matrix up to 84% on the largest. However, after 50% reduction, solving the remainder takes only a fraction of the time required to solve the entire system, with the savings ranging from 81–97%.

In Table 6, we show the results for combined parallel processing and partial system reduction on the circuit-based matrices. Once again, only the results after 50% reduction are presented as these have a greater impact on parallel performance.

Table 6. Recorded times for the partial GJE algorithm with 50% of columns reduced using serial and parallel execution on circuit-based matrices

| Matrix Size | 1 thread | 2 threads | 4 threads | 8 threads |
|---|---|---|---|---|
| 128×129 | 38.4 μs | 29.8 μs | 38.3 μs | 46.8 μs |
| 512×513 | 544.9 μs | 511.8 μs | 515.8 μs | 576.7 μs |
| 1024×1025 | 3.2 ms | 2.6 ms | 2.5 ms | 2.3 ms |
| 2040×2041 | 33.9 ms | 13.7 ms | 11.6 ms | 9.5 ms |

From this table, one can observe inferior performance from the continued increase in parallelisation on the two smaller matrices, as two threads are optimal for both. The $1024 \times 1025$ matrix records lower parallel efficiency following the partial reduction ($1.4 \times$ speedup on eight threads) though parallelism is faster. However, the largest matrix sees an improvement in speedup, with up $3.6 \times$ on eight threads.

*4.3. Comparison of Gauss-Jordan Elimination on Random and Circuit Matrices*

Overall, our implementation of the GJE algorithms yields notably quicker execution times when solving the sparser systems discussed in Section 4.2 compared to the denser ones in Section 4.1. However, parallel performance suffers likely due to an uneven split of work between the processors (compare Table 1 and Table 4). Solving the remainder of the system after it has been partially reduced is much quicker on the sparser systems than on the denser ones, with a notably greater percentage of time saved (compare Table 2 and Table 5). In both cases, larger matrices benefit more than smaller ones. Combining both methods provides mixed results. Parallel performance is generally maintained on medium-size matrices, with only smaller sparse ones seeing a reduction in efficiency, whilst the largest matrices see an improvement in parallel performance (compare Table 3 and Table 6).

## 5. Conclusion

In this paper, we considered an iterative circuit simulation model that relies on solving a system of linear equations in each iteration by Gauss-Jordan elimination (GJE). We then proposed two innovative approaches to improve our model's performance: parallel programming for GJE and partial system reduction.

The proposed parallel GJE algorithm is designed to partition the matrix among the processors in a row-wise fashion as uniformly as possible. We also introduced a separate "pivot matrix" for this algorithm to store the pivot rows so as to minimise synchronisation between threads throughout the process. Partial system reduction was proposed to simplify the system prior to initiating iterative simulation. Our method considers time-varying entries and respects constraints that ensure that inserting their values at different moments in time remains valid. The theoretical analysis of this approach implied a quadratic reduction in time when the remainder of the system is still to be solved at each iteration. The two methods are also mutually independent and can be used in combination.

As such, we tested the two approaches both in isolation and in combination on dense and sparse linear systems. Parallel GJE alone provided a notable speedup across all input sizes. An increase in speedup was recorded as the number of processors increased, but efficiency was not maintained. A greater parallel speedup was observed on denser systems than on sparser ones. After partial system reduction, solving the remainder of the system was notably quicker, as the theoretical analysis had implied. Sparser systems saw a greater reduction in time than denser ones. For most matrix sizes and types, after partial system reduction, parallel GJE maintained a notable speedup and recorded comparable behaviour to earlier tests. On the smaller sparse matrices, efficiency was significantly reduced to the point that in some cases, parallel GJE was slower than the sequential equivalent. On the other hand, the largest matrices saw an increase in speedup after the system was simplified.

As future work, we would like to explore the performance of our approach when simulating electric circuits. From here, we aim to develop a simulator for educational purposes. Our aim is to support both time-dependent components, as described earlier in this paper, and reactive components such as capacitors and inductors, by exploring an approach utilising Thévenin's theorem. Here, we would replace the analytical approach with one that involves solving a modified version of the system of equations corresponding to a given electric circuit.

**Declaration of Competing Interests**

There are no competing interests.

**Acknowledgements**

The authors would like to thank their colleague Atanas Hristov for his valuable advice and support during the preparation of this paper.